\documentclass[11pt,a4paper]{article}

\usepackage[utf8]{inputenc}
\usepackage[T1]{fontenc}
\usepackage{lmodern}
\usepackage{amsmath,amssymb}
\usepackage{graphicx}
\usepackage{float}
\usepackage{hyperref}
\hypersetup{colorlinks=false}
\usepackage{booktabs}
\usepackage{caption}
\usepackage{subcaption}
\usepackage{geometry}
\usepackage{authblk}

\begin{document}

% Title
\title{Impact of Residual Angular Chirp in a Petawatt-class Laser System on Laser-driven Proton Acceleration}

% Authors
\author{Qingfan Wu$^{1}$, Minjian Wu$^{1}$, Jiarui Zhao$^{1,a}$, Ying Gao$^{1}$, Haoran Chen$^{1}$, Tan Song$^{1}$, Zhongshuai Zhang$^{1}$, Zhangyi Wu$^{1}$, Tianhao Liang$^{1}$, Shirui Xu$^{1}$, Ziyang Peng$^{1}$, Hui Zhang$^{1}$, Tianqi Xu$^{1}$, Qihang Han$^{1}$, Chenghao Hua$^{1}$, Ke Chen$^{1}$, Pengcheng Fan$^{1}$, Yuntian Xie$^{1}$, Xianduo Li$^{3}$, Peiqiang Liu$^{3}$, Xiangyu Nong$^{3}$, Shengxuan Xu$^{1}$, Liyong Ma$^{1}$, Yixing Geng$^{1}$, Chen Lin$^{1,2,3}$, Yanying Zhao$^{1,2,3}$, Xueqing Yan$^{1,2,3,4}$ and Wenjun Ma$^{1,2,3,4,a}$}

\affil{$^{1}$State Key Laboratory of Nuclear Physics and Technology, School of Physics, Peking University, Beijing 100871, China}
\affil{$^{2}$Beijing Laser Acceleration Innovation Center, Huairou, Beijing 101400, China}
\affil{$^{3}$Institute of Guangdong Laser Plasma Technology, Baiyun, Guangzhou 510540, China}
\affil{$^{4}$Collaborative Innovation Center of Extreme Optics, Shanxi University, Taiyuan, Shanxi 030006, China}

\affil{$^{a}$) Authors to whom correspondence should be addressed: wenjun.ma@pku.edu.cn, jrzhao@pku.edu.cn}

\date{}

\maketitle

% Abstract
\begin{abstract}
Laser-driven proton acceleration has attracted considerable interest owing to its appealing potential in versatile applications including cancer therapy. Proton energies depend critically on the on-target intensities, yet the detrimental impact of focal spot degradation induced by spatiotemporal couplings on the acceleration remains insufficiently elucidated. In this study, we demonstrate that residual angular chirp (AC), stemming from minor misalignments of the grating compressor in a Petawatt-class laser system, acts as a critical bottleneck for proton acceleration. Experimental results reveal that even around 100 microradians of grating misalignment induces substantial focal-spot elongation and a pronounced reduction in peak intensity. By implementing an in situ spectral-blocking diagnostic, we effectively eliminated the residual AC and restored a near-diffraction-limited focus. This optimization led to a significant recovery of the on-target intensity, resulting in a twofold increase in the proton cutoff energy. Our work presents a successful demonstration of diagnosing and eliminating residual AC. This provides a practical reference for generating high-energy proton beams and supporting their diverse applications in a PW-class laser.
\end{abstract}

% Keywords

\section{Introduction}\label{introduction}

Laser-driven proton acceleration has advanced rapidly over the past two decades\cite{1,2,3}. The resulting proton beams, characterized by sub-picosecond durations, micrometer-scale source sizes, and high peak fluxes\cite{4,5}, hold transformative potential for cancer radiotherapy\cite{6}, inertial confinement fusion\cite{7}, proton imaging\cite{8,9,10}, and warm dense matter studies\cite{11,12}. To date, significant experimental efforts have been dedicated to pushing proton energies beyond 100 MeV via various acceleration mechanisms such as Target Normal Sheath Acceleration (TNSA)\cite{13}, Radiation Pressure Acceleration (RPA)\cite{14,15}, and Relativistic Induced Transparency (RIT)\cite{15,16}. Among all acceleration regimes, the energy and flux of the protons are highly sensitive to the laser's peak intensity. While a growing number of world-class facilities have demonstrated the capability to reach peak intensities exceeding \(10^{21}\)W/cm$^2$\cite{17,18},  ensuring the consistent and reproducible delivery of such peak performance remains a universal challenge.

Realizing extreme peak intensity with PW-class lasers requires near-perfect control of both spatial and temporal pulse characteristics. Standard techniques include precise dispersion management for pulse compression and deformable mirror (DM) to correct monochromatic wavefront aberrations\cite{19,20,21}. Nevertheless, even with a perfectly flat wavefront and an optimized focal spot, the peak intensity can be significantly degraded by Spatiotemporal Couplings (STCs)\cite{22,23}. STCs refer to the phenomenon where different spectral components of the laser pulse exhibit spatial variations in intensity or wavefront, resulting in spatially dependent temporal characteristics (e.g., spatial chirp and pulse front tilt)\cite{24,25}. Such effects are particularly severe in PW laser systems owing to their large-aperture beams and broad spectral bandwidths.

Angular chirp (AC) is one of the most noteworthy STCs in PW laser systems\cite{26}. In particular, generating PW laser pulses shorter than 30 fs requires large-aperture diffraction gratings in the compressors. Such large gratings render the compressor highly sensitive to alignment errors, even a slight misalignment of the gratings is sufficient to induce residual AC\cite{27}. This residual AC impairs the spatiotemporal concentration of energy via spectral-angle coupling. To date, various diagnostic tools such as IFA\cite{26}, SRI\cite{28}, and GRENOUILLE\cite{25}, have been developed to characterize AC, typically by relying on sampled, collimated beams. In addition, several complex methods have been proposed to diagnose residual STCs by characterizing the near-field wavefronts of full-aperture beams across different spectral components\cite{29,30,31,32}. However, the true impact of AC is manifested at the focus. Specifically, when a beam carrying AC is focused by an off-axis parabolic mirror (OAP), the distinct spectral components deviate from an ideal spatial overlap at the focal plane\cite{26}. This results in an elongated focal spot---a phenomenon termed `spatial chirp at focus'---along with localized pulse broadening. Consequently, although the total laser energy remains conserved, the effective peak intensity is dramatically "diluted".

Pretzler et al. established a comprehensive quantitative framework relating AC to key parameters such as focal spot broadening, pulse stretching, and specific grating properties\cite{26}. Building on this foundation, Zeil et al. demonstrated that introducing controlled AC by precisely adjusting compressor grating angles can effectively steer proton beam emission\cite{33}. Furthermore, Grace et al. utilized 2D simulations to show that pulse front tilt (a direct consequence of residual AC) can be exploited to manipulate hot electron trajectories and subsequently reshape the sheath electric field distribution.\cite{34} However, these studies have primarily employed AC to control proton beam pointing or to steer hot electron trajectories within the plasma. As the number of PW-class facilities worldwide continues to grow\cite{35}, it is essential to systematically investigate both the practical limitations that residual AC places on proton acceleration, and the quantitative operation bounds and energy scaling relationships needed for performance optimization.\\
In this work, we integrate established knowledge on AC characterization and mitigation into a quantitative experimental study at the CLAPA-II 2 PW laser system. We demonstrate that residual AC constitutes a significant yet often overlooked limitation in PW-class laser-proton acceleration. Specifically, AC-induced focal spot degradation persists despite near-diffraction-limited wavefront quality. This coupling imposes an upper bound on the peak intensity, thereby constraining the maximum proton energy. By employing a simple spectral-blocking diagnostic to characterize the focal-plane distribution of different spectral components, we directly identified the residual AC at the interaction point and quantified its impact on the peak intensity. Through precise fine-tuning the angles of the compressor gratings, we eliminated the residual AC and obtained a near-ideal focal spot, thereby boosting the maximum proton energy by more than a factor of two.

\section{Experiment setup}\label{experiment-setup}

\begin{figure}[H]
\centering
\includegraphics[width=\textwidth]{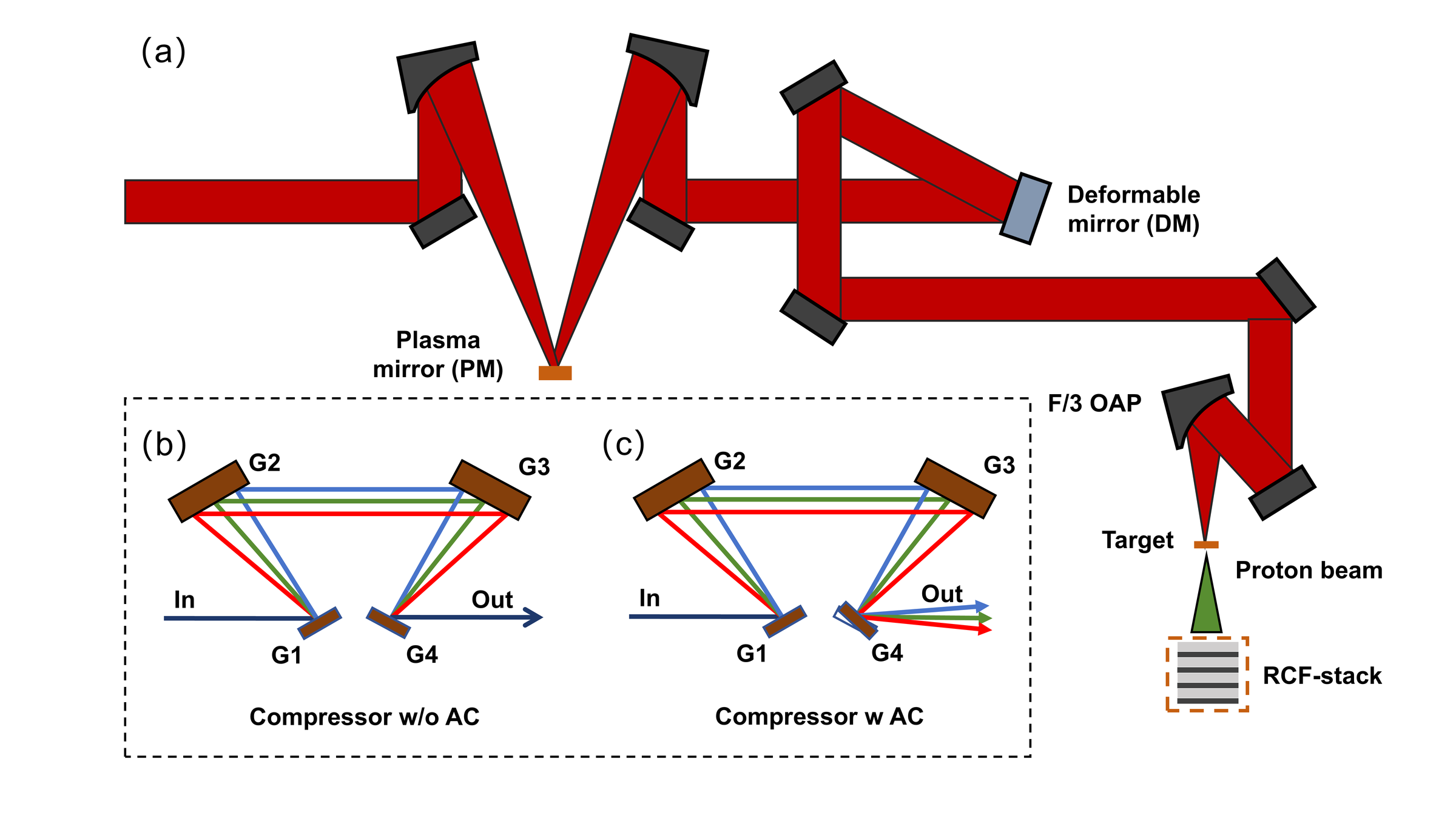}
\caption{(a) Beamline in the experimental area. (b) Schematic of a typical compressor without AC. (c) AC induced due to the misalignment of a single compressor.}
\label{fig:1}
\end{figure}

The experiments were conducted using the CLAPA-II facility\cite{36,37,38}, a 2 PW Ti:sapphire laser system with a maximum pulse energy of 60 J. After the final amplification stage, the beam enters a standard four-grating compressor (groove density: 1480 lines/mm), where it is compressed to a transform-limited duration of approximately 30 fs. To ensure that proton acceleration was governed by the main pulse interaction and to suppress pre-plasma formation that might obscure the effects of AC, the intrinsic laser contrast of \({3.8 \times 10}^{- 11}\ \)(measured in target area at -100 ps) was further enhanced by approximately two orders of magnitude through a plasma mirror (PM) system [see Fig.~\ref{fig:1}(a)]. A periscope chamber for the conversion of the polarization was inserted between the compressor and the PM system to convert the polarization of the pulse from P to S. The PM system consisted of an anti-reflection-coated glass substrate (reflectivity $<$ 0.1\%) irradiated with an \emph{s}-polarized pulse at a fluence of \(200\ J/{cm}^{2}\) (yielding a plasma reflectivity of $\sim$80\%) and two OAPs with F-numbers F/3.5 and F/2.8, which were used to recollimate the beam, producing a 300 mm output beam.

After the PM system, the beam was directed to a large-aperture DM (500 mm diameter, 129 actuators, ISP System, France) to correct cumulative wavefront aberrations. Finally, the beam was focused onto the target using an F/3 OAP (effective focal length: 990 mm, off-axis angle: 45$^\circ$) at an incidence angle of 5$^\circ$ in an s-polarized geometry, which, in the absence of AC, is capable of delivering a near-diffraction-limited focal spot of \(3.4 \times 3.2\ \mu m\) (full-width at half-maximum, FWHM). The targets used in the experiment were polymer foils with a thickness of 50 nm. The accelerated protons were characterized along the laser axis using stacks of radiochromic film (RCF).

In a high-power femtosecond laser system, residual AC mainly arises from the misalignment of large-aperture compressor gratings\cite{27,39}. Fig.~\ref{fig:1}(b) shows a schematic of typical compressor in a PW laser system. The compressor comprised four gratings (G1--G4), leveraging angular dispersion to introduce temporal dispersion for pulse compression. Even a subtle rotational misalignment of a single compressor grating can readily introduce residual AC into the output beam [see Fig.~\ref{fig:1}(c)]. As a result, different spectral components acquire distinct angular deviations, preventing them from achieving an ideal spatial overlap at the focus. In the following sections, we describe the in-situ diagnostic used to identify these misalignments and the subsequent optimization process employed to eliminate them.

\section{Experiment Results}\label{experiment-results}

\subsection{Wavefront Optimization}

\begin{figure}[H]
\centering
\includegraphics[width=\textwidth]{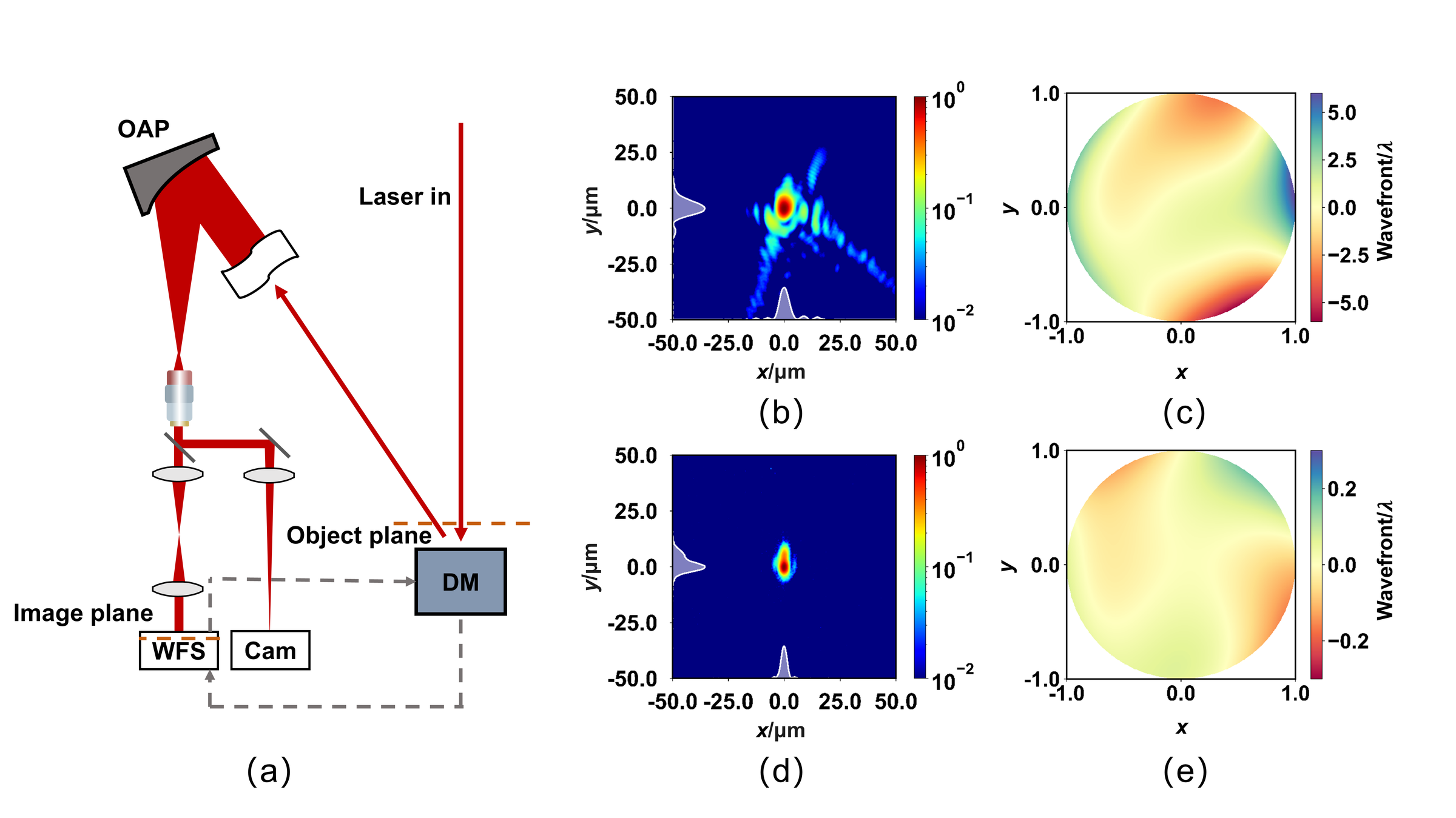}
\caption{(a) Layout of the AO system. (b) Spot and (c) phase before wavefront correction. (d) Spot and (e) phase after wavefront correction.}
\label{fig:2}
\end{figure}

To compensate for the cumulative wavefront aberrations across the optical chain, we first employed an adaptive optics (AO) system consisting of a wavefront sensor (Phasics, SID4-V) and a DM to perform closed-loop aberration correction at a pulse energy of 1 mJ. As illustrated in Fig.~\ref{fig:2} (a), the laser beam is reflected by the DM and focused by an F/3 OAP in the target chamber. Subsequently, the real-time wavefront was measured by a wavefront sensor (Phasics, Sid4-V) operating at 1Hz. The wavefront sensor was mounted in an imaging system consisting of an OAP, a microscope objective, and two lenses. Its image plane is located near the DM plane. Based on real-time wavefront sensing, the AO system iteratively optimized and corrected the wavefront toward a perfect planar wavefront, minimizing the near-field wavefront aberrations. Meanwhile, to measure the far-field focal spot online, an imaging lens with a focal length 30 times that of the microscope objective was inserted into the collimated beam path downstream. The far-field intensity distribution was captured by a CCD (GED200M, Mindvision). This configuration yielded a 30$\times$ magnification of the focal spot, allowing for the real-time characterization of its morphology concurrently with the wavefront optimization.

Initial measurements Fig.~\ref{fig:2} (b) and (c) revealed that the uncorrected beam contained severe wavefront distortions with a peak-to-valley (PV) of 4.1\(\lambda\), a root-mean-square (RMS) of 0.85\(\lambda\), and a Strehl ratio (SR) of only 0.048. The focal spot was fragmented, exhibiting full-widths at half-maximum (FWHM) of 4.99 $\mu$m (X) and 7.37 $\mu$m (Y) and an energy concentration within the FWHM of 18\%. After the closed-loop optimization [Fig.~\ref{fig:2} (d) and (e)], the wavefront quality was improved markedly, yielding a PV of 0.453\(\lambda\), an RMS of 0.049\(\lambda\), and an SR of 0.91. Under such conditions, a near-diffraction-limited focal spot is anticipated. Notably, the measured focal spot remained notably elongated into an elliptical shape, with FWHM dimensions of 3.10 $\mu$m and 5.88 $\mu$m along the X and Y directions, respectively. The energy concentration within the FWHM was 36\%. This discrepancy between the high-quality wavefront and the degraded focal spot morphology strongly implies that the beam's focusing ability is not limited solely by spatial phase errors. This suggests the presence of frequency-angle coupling, specifically residual angular chirp (AC), which is undetectable by standard monochromatic wavefront sensors yet induces severe attenuation of peak intensity at the focal plane.

\subsection{Diagnostic and elimination of Angular Chirp}

\begin{figure}[H]
\centering
\includegraphics[width=\textwidth]{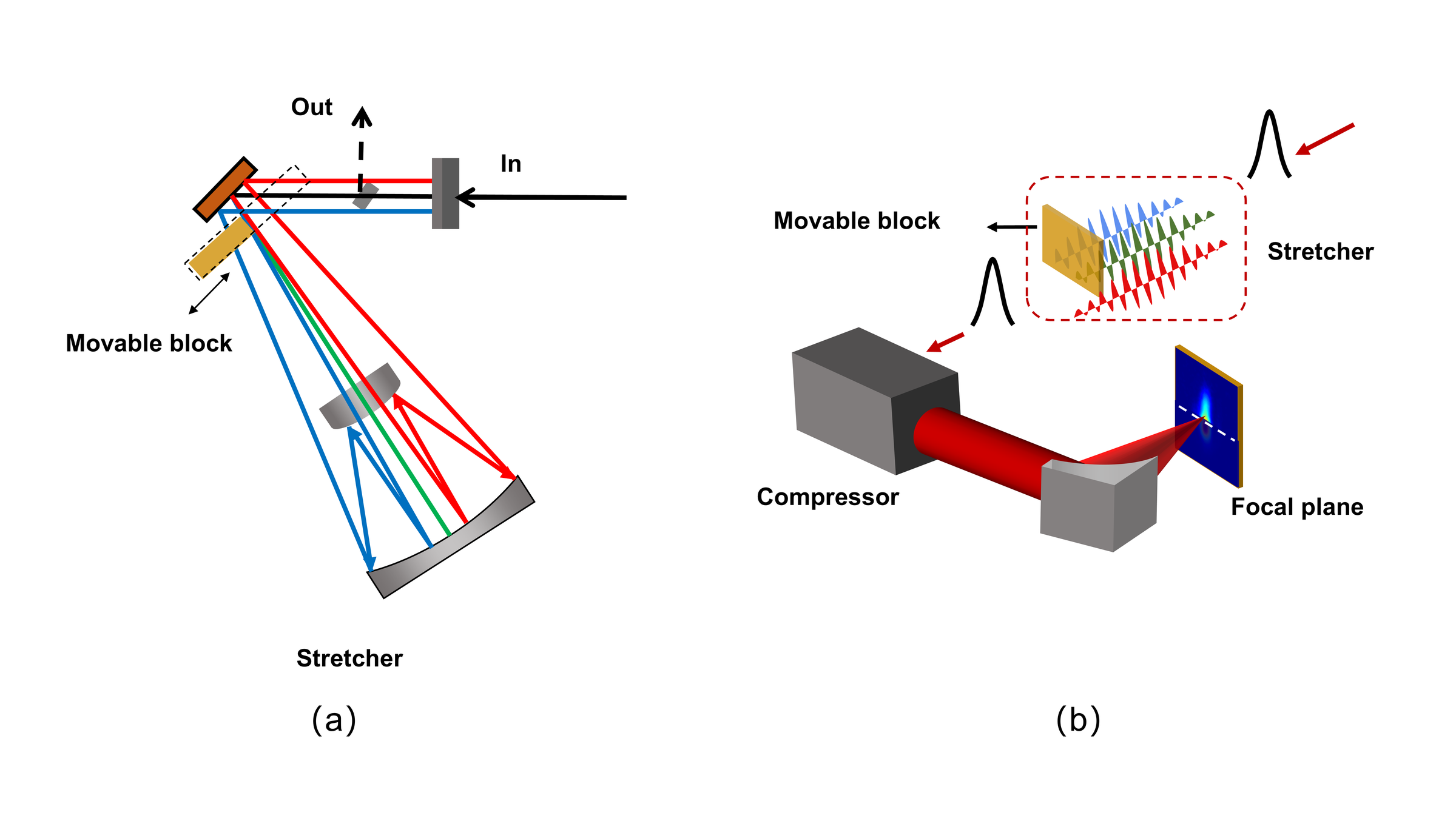}
\caption{(a) Layout of the stretcher; (b) Schematic of spectral-blocking focus diagnostic.}
\label{fig:3}
\end{figure}
To identify the origin of the observed focal spot elongation, we implemented an in-situ diagnostic by using a spectral-blocking setup in the stretcher, building on the concept of spectrally-resolved measurement\cite{29,40}. As illustrated in the stretcher schematic [Fig.~\ref{fig:3} (a)], the pulse is spatially dispersed by a grating and reflected by concave and convex mirrors, resulting in a spatially resolved spectral line profile on the grating surface upon return. Within this line profile, each transverse position corresponds to a specific spectral component. By placing a movable blocker before the grating surface, we can selectively gate either the short-wavelength or long-wavelength wings of the pulse. These isolated spectral segments are subsequently compressed and focused by an OAP onto the target [Fig.~\ref{fig:3} (b)]. By imaging the focal spots on the far-field CCD, we can directly compare the intensity distributions of different spectral components. This method exploits the fact that spectrally separating the beam and observing the resulting focal-spot distribution provides an effective means of diagnosing spatiotemporal couplings. It overcomes the inherent limitations of standard monochromatic wavefront sensors regarding spatiotemporal-coupling detection.

\begin{figure}[H]
\centering
\includegraphics[width=\textwidth]{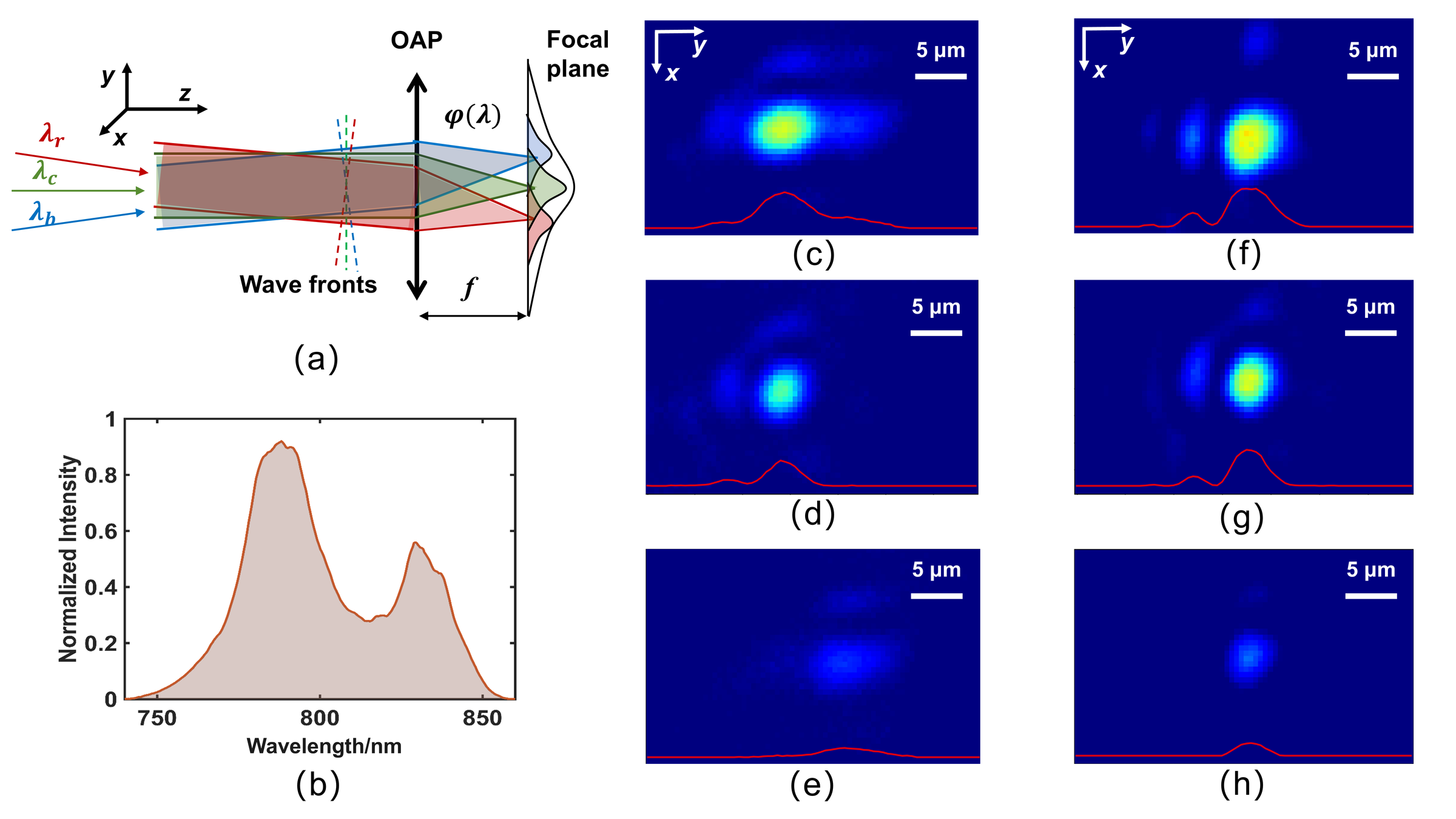}
\caption{(a) Schematic of spatial-spectral coupling. (b) Spectral measurement at 1 mJ. (c) Elongated focal spot (full spectrum) with residual AC. (d) Long-wavelength-blocked focal spot with residual AC. (e) Short-wavelength-blocked focal spot with residual AC. (f) Optimized focal spot (full spectrum) after AC elimination. (g) Long-wavelength-blocked focal spot after AC elimination. (h) Short-wavelength-blocked focal spot after AC elimination, showing consistent spot morphology.}
\label{fig:4}
\end{figure}

Consider a pulse propagating in the z direction possessing residual AC in the y-z plane [Fig.~\ref{fig:4} (a)]. Under these conditions, different spectral components exhibit distinct propagation directions, analogous to the effect of a dispersive element. When such an AC-affected beam is tightly focused, this angular dispersion leads to spectral decomposition, where the transverse position of each frequency component is determined by its specific propagation angle. Consequently, the intrinsic spectral profile of the pulse is directly mapped onto the spatial intensity distribution of the focal spot. We measured the power spectrum of the pulse using a high-resolution spectrometer (HR4000, Oceanview) and found a higher intensity in the short-wavelength region [Fig.~\ref{fig:4}(b)], a spectral non-uniformity that is highly consistent with the asymmetric transverse intensity distribution of the unblocked, elongated focal spot [Fig.~\ref{fig:4}(c)]. To confirm this interpretation, we blocked the 800--850 nm and 750--800 nm spectral components in sequence. Fig.~\ref{fig:4}(d) and 4(e) show that blocking the 800--850 nm components extinguished the right half of the focal spot, whereas blocking the 750--800 nm components extinguished the left half. This spatial-spectral mapping constitutes definitive evidence of residual AC.

To eliminate the residual AC, we iteratively adjusted the orientation angle of the last grating of the compressor, using the symmetry of the focal spot as the primary optimization metric. The effectiveness of this correction is evident in the comparison between the "before" and "after" states. As shown in the optimized results [Fig.~\ref{fig:4}(f)--(h)], the focal spots of different spectral components---previously separated---now converged onto the same transverse coordinates. This "re-stacking" of the spectral components in the spatial domain confirms the elimination of the residual AC.

\begin{figure}[H]
\centering
\includegraphics[width=0.7\textwidth]{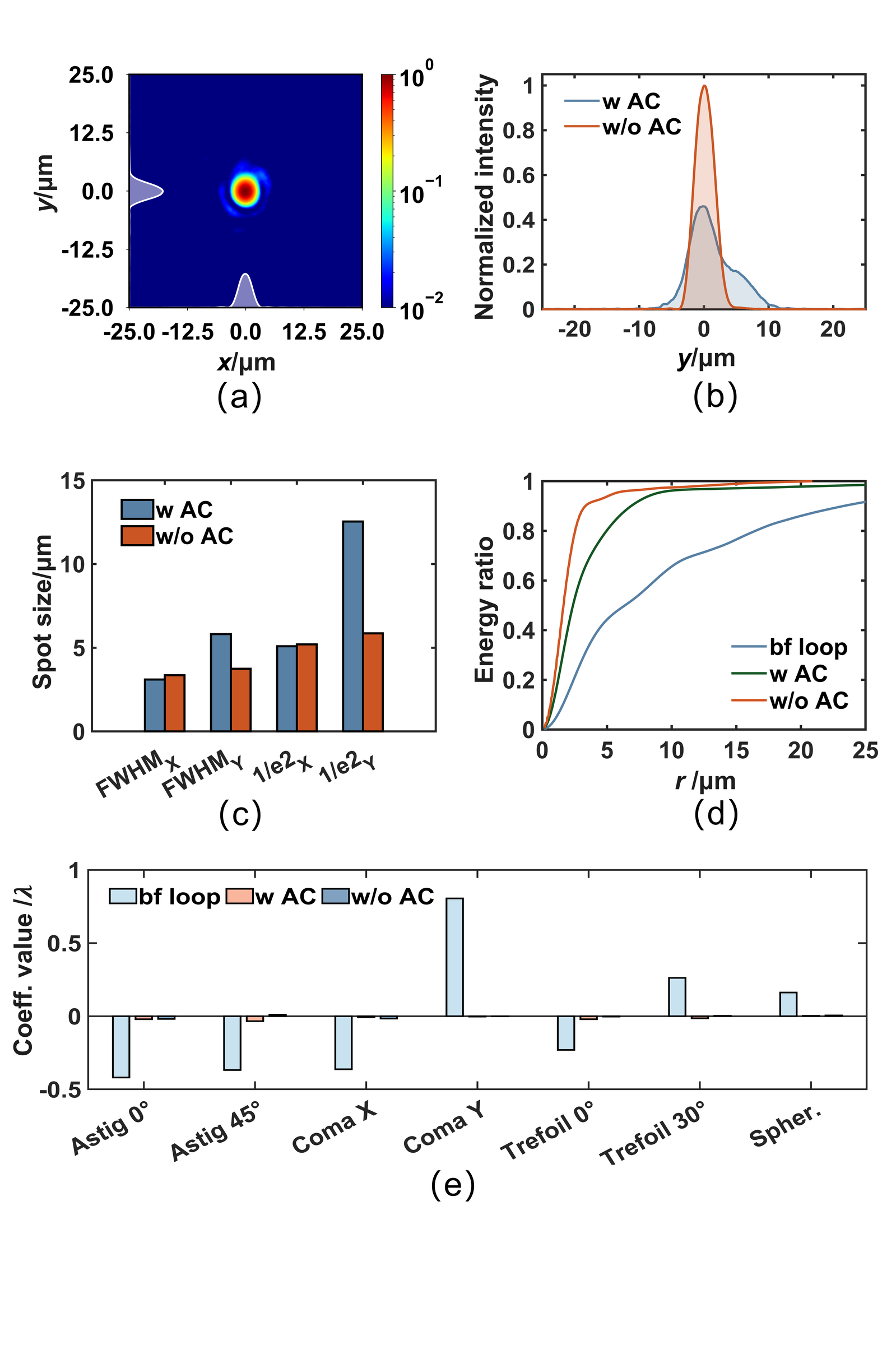}
\caption{Comparison of the laser's parameters before and after AC elimination. (a) Near-diffraction-limited focal spot without AC. (b) Intensity profiles, (c) Focal spot size variation, (d) Encircled energy fraction, and (e) Zernike polynomial coefficients before and after AC elimination. ``bf loop'' denotes the state before the adaptive optics closed-loop correction.}
\label{fig:5}
\end{figure}

After the elimination of residual AC, wavefront correction was implemented again. As illustrated in Fig.~\ref{fig:5}(a), a near-field wavefront (SR=0.91) and a near-diffraction-limited focal spot were obtained. Notably, the corrected focal spot shown in Fig.~\ref{fig:5}(a) exhibited a FWHM of 3.54$\times$3.34 $\mu$m. Compared to the focal spot elongation along the y-axis caused by residual AC [Fig.~\ref{fig:5}(b)], the optimization achieved by AC elimination resulted in a much more symmetric and concentrated intensity distribution. Specifically, the spot size with AC increased by a factor of 2 compared to the case without AC, based on the \(1/e^{2}\) intensity level [Fig.~\ref{fig:5}(c)]. The calculated encircled energy curves further validated the improved energy concentration after elimination [Fig.~\ref{fig:5}(d)], with the spatial intensity increasing nearly two-fold. Crucially, the measured wavefront remained unchanged throughout this process, ruling out spatial wavefront aberrations and confirming that initial focal elongation stemmed exclusively from spectral separation [Fig.~\ref{fig:5}(e)].

To evaluate the overall impact of residual AC on the peak intensity, the corresponding temporal stretching must also be taken into account. While direct measurement of the local pulse duration at the tight focus of a PW-class laser remains a significant diagnostic challenge, the intensity enhancement can be theoretically estimated. In our analysis, we employ the established analytical formalism for angular chirp and grating compressor misalignment\cite{24}. Specifically, AC not only spatially broadens the focal spot but also temporally stretches the pulse duration, as the spectral width is reduced at each point of the focus. Because the constituent spectral components remain spatially overlapped in the collimated beam before focusing, this pulse broadening at the focal spot cannot be detected by conventional collimated-beam diagnostics. The input parameters for this analytical model were directly extracted from our in-situ 30$\times$ magnified focal plane images Fig.~\ref{fig:2} (d). 

According to Equations (1) and (2), the pulse width stretching factor is approximately equal to the spatial broadening factor. Here, \(\xi\) denotes the pulse width (or spot size) broadening factor. \(\omega_{AC}\) and\(\ \omega_{0}\) represent the radius of the first Airy disk zero with and without AC, respectively; Their ratio is approximately equal to that of the corresponding 1/e$^2$ beam radii. \(\Delta\lambda\) is the spectral bandwidth, \(f\ \)is the focal length, \(\tau_{AC}\) and\(\ \tau_{0}\) are the pulse width with and without AC, and \(C_{a}\) is the angular chirp value (\(\frac{d\varphi}{d\lambda}\)). 

Accordingly, our calculations yield the peak intensities before and after AC elimination as: \({1.10 \times 10}^{20}\ \)W/cm$^2$ and\({\ 5.01 \times 10}^{20}\) W/cm$^2$, respectively. Furthermore, by substituting the spatial broadening factor of \(\xi = 2.1\) into Equation (2) along with other laser parameters (\(\omega_{0} = 3.22\ \mu m,\ \omega_{AC} = 6.75\ \mu m,\) \(\Delta\lambda = 50nm\), \(f = 0.99\ m\)), the residual AC was quantitatively determined to be \(C_{a} = 0.14\ \mu rad/nm\). Based on Equation (3) for our conventional four-grating compressor with only the fourth grating misaligned, the corresponding grating misalignment angle (\(\varepsilon_{x}\)) is estimated to be approximately 137 $\mu$rad.

\begin{equation}
\xi = \frac{\omega_{\mathrm{AC}}}{\omega_{0}} = \frac{\tau_{\mathrm{AC}}}{\tau_{0}}
\label{eq:1}
\end{equation}

\begin{equation}
\xi = \sqrt{1 + \left( \frac{C_{a}f\delta\lambda}{\omega_{0}} \right)^{2}} = \sqrt{1 + \left( \frac{C_{a}f\Delta\lambda}{\sqrt{2\ln 2}\omega_{0}} \right)^{2}}
\label{eq:2}
\end{equation}

\begin{equation}
C_{a} = \varepsilon_{x}N\frac{\tan\beta_{0}}{\cos\alpha}
\label{eq:3}
\end{equation}

In Equation (3), which relates angular chirp to grating misalignment in our four-grating Treacy compressor, N=1480 mm$^{-1}$ is the groove density, while \(\alpha\)=53$^\circ$ and \(\beta_{0}\)=22.6$^\circ$ are the incident and diffraction angles at the central wavelength, respectively.

\subsection{Proton Acceleration Results}

To evaluate the influence of residual AC on the efficacy of laser-driven proton acceleration, we conducted comparative experiments using 50-nm-thick polymer foils at the CLAPA-II facility at a laser power of 100 TW, utilizing the PM system to ensure high temporal contrast. The proton energy spectra and spatial distributions were characterized using RCF stacks, both before and after the elimination of AC. The hybrid RCF stack (purchased from Ashland) was composed of four layers of HD-V2 films for high-flux low-energy protons, followed by four layers of EBT3 films for high-energy components.

\begin{figure}[H]
\centering
\includegraphics[width=\textwidth]{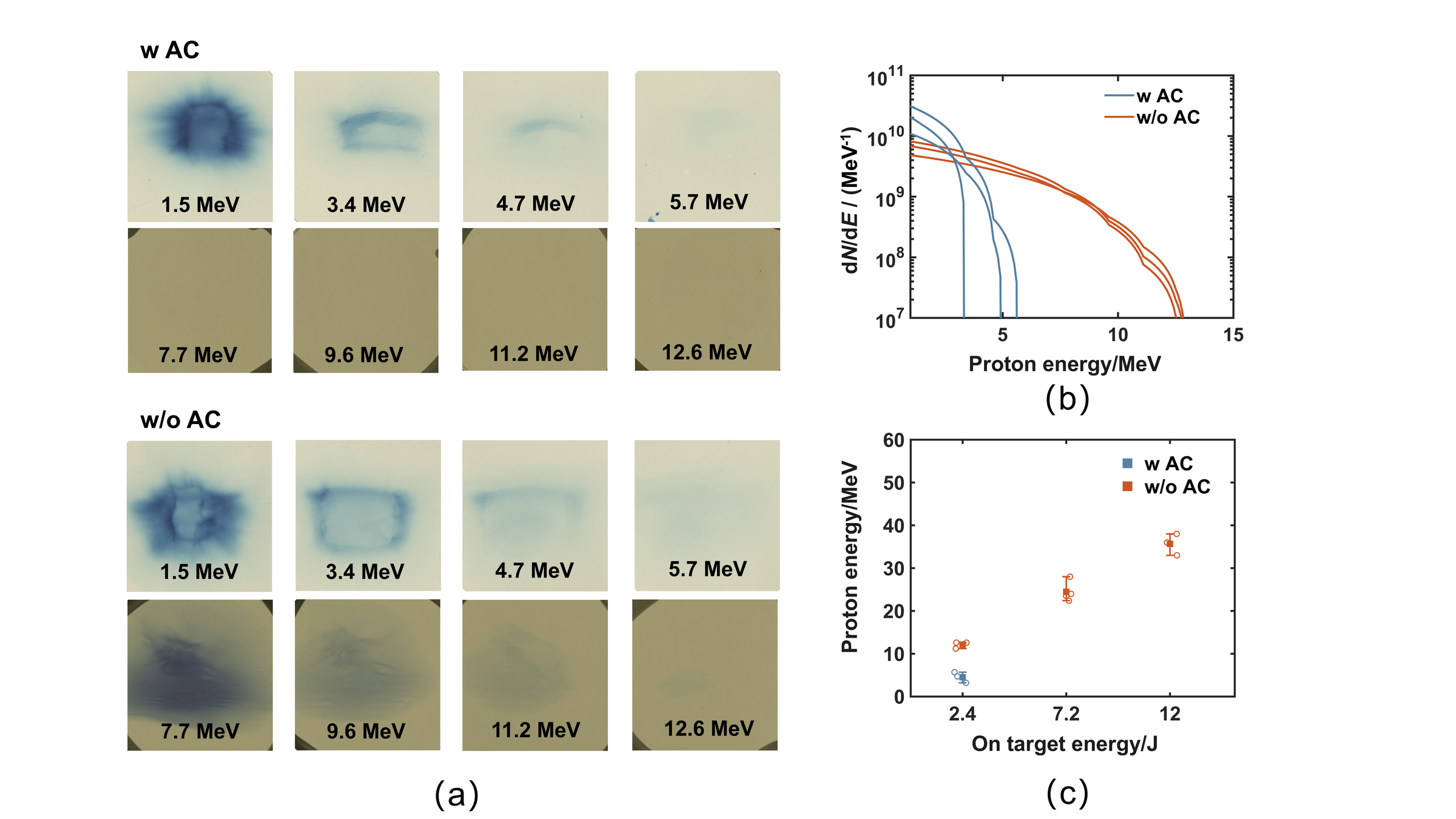}
\caption{Enhancement of laser-driven proton acceleration via AC elimination. (a) RCF measurements. (b) Reconstructed proton energy spectra from the RCF data before and after the AC elimination. (c) Proton cut-off energies at distinct laser powers. Power values (100, 300, 500 TW) correspond to 2.4, 7.2, 12 J on target after the $\sim$20\% PM insertion loss. Error bars indicate the shot-to-shot variation range under identical experimental conditions.}
\label{fig:6}
\end{figure}

The experimental results, as summarized in Fig.~\ref{fig:6}, revealed a profound enhancement in acceleration performance after the elimination of residual AC. Before optimization, the maximum proton energy was about 5.7 MeV. As shown in Fig.~\ref{fig:6} (a), the RCF images showed low proton flux, consistent with reduced laser intensity. Following AC elimination and restoration of focal-spot quality, the proton cutoff energy increased markedly to 12.6\,MeV---a twofold increase. The corresponding multi-shot energy spectra reconstructed from the RCF data are compared in Fig.~\ref{fig:6} (b). In addition to the enhanced peak energy, the proton beam exhibited a significantly higher flux and improved symmetry in spatial profile [see Fig.~\ref{fig:6} (a)], with the total number of protons above 1.5 MeV increasing from \({\ 1.81 \times 10}^{10}\) under the influence of residual AC to \({\ 2.80 \times 10}^{10}\) after its removal. This improvement indicates that the restoration of spatiotemporal quality at the focus is essential for accessing the high-intensity regime required for efficient acceleration.

The enhancement in proton energy is consistent with the TNSA mechanism, which governs the interaction under the present 100-TW laser and target parameters. In the TNSA regime, the maximum proton energy \(E_{\max}\) is known to scale as the square root of the peak laser intensity (\(E_{\max} \propto \sqrt{I}\)). Our focal spot characterization in Section 3.2 indicated that the elimination of AC led to a nearly fourfold increase in peak intensity. According to the scaling law, a fourfold increase in intensity should double the maximum proton energy. Our experimental findings, showing an energy increase from 5.7 MeV to 12.6 MeV (a factor of 2.2), agree well with this prediction. Physically, the elimination of AC provides a dual effect: it simultaneously boosts the peak intensity and shortens the local pulse duration. While a shorter pulse might limit the acceleration time, the strong agreement with the \(\sqrt{I}\) scaling suggests that the enhancement of hot electron temperature remains the dominant driver here.

To assess the performance of the optimized laser beam beyond the 100 TW benchmark, we performed a laser energy scan. The resulting proton cut-off energies are shown in Fig.~\ref{fig:6} (c). It is clear that after the elimination of residual AC, the system exhibits stable scale up of the cut-off energies with increasing laser energy, showing a better-than-\(\sqrt{I}\) scaling trend. Similar trends have been reported in short-pulse laser-driven proton acceleration\cite{41}. Specifically, the average cut-off energy reaches 24.5 MeV at a laser power of 300 TW, and further increases to an average of 35.6 MeV as the power is scaled up to 500 TW, demonstrating the robust performance of the optimized system. Here, the laser power values represent the nominal output after the grating compressor, which translate to net on-target energies of 2.4 J, 7.2 J, and 12 J after accounting for the $\sim$20\% insertion loss of the PM system.

\section{Conclusion}\label{conclusion}

In summary, we have shown that residual angular chirp is a primary, yet often hidden, factor limiting proton acceleration in PW-class lasers. Even a 137 $\mu$rad alignment errors in the compressor can severely degrade the peak intensity, even when the spatial wavefront appears nearly perfect. By employing a simple, in-situ spectral-blocking diagnostic, we directly captured the spatial-spectral separation at the focus and restored its spatiotemporal integrity. This optimization led to a twofold increase in proton cutoff energy under 100-TW conditions. The result is in excellent agreement with the TNSA scaling law (\(E_{\max} \propto \sqrt{I}\)), confirming that the elimination of AC effectively quadrupled the on-target peak intensity.

Our findings demonstrate that as laser facilities approach the 10-PW scale, managing these spatiotemporal couplings will be just as critical as spatial wavefront control. Standard adaptive optics alone is no longer sufficient to guarantee peak experimental performance. The diagnostic and mitigation strategies presented here are essential for ensuring the optimal delivery of extreme intensities, which is critical for exploring new frontiers in high-field physics and advanced laser-ion applications. Future work addressing the interplay between AC and other spatiotemporal couplings\cite{32,42,43}, such as longitudinal chromatic aberration, as well as the diagnosis and optimization of laser focal spots under high-power conditions\cite{44,45,46}, will be essential to further push the limits of laser-matter interactions at the highest intensities.

\section*{Acknowledgement} This work was supported by the following projects: the National Key Research and Development Program of China (Grant No. 2024YFF0726304), National Science Fund for Distinguished Young Scholars (Grant No. 12225501), the National Grand Instrument Project (Grant No. 2019YFF01014402), the Beijing Natural Science Foundation (Grant No. 1252019), the National Natural Science Foundation of China (Grant Nos. 12575257, 12595360, 12595361, 12595362, 12205008), the Guangdong High Level Innovation Research Institute (Grant No. 2021B0909050006). Wavefront data analysis was performed on the High-Performance Computing Platform of Peking University.

\section*{Author Declarations}
\textbf{Conflict of Interest}

The authors have no conflicts to disclose.

\section*{Data Availability} The data that support the findings of this study are available from the corresponding author upon reasonable request.

\end{document}